\RequirePackage{fix-cm}
\documentclass[letterpaper, 10 pt, conference]{ieeeconf}

\IEEEoverridecommandlockouts
\def\BibTeX{{\rm B\kern-.05em{\sc i\kern-.025em b}\kern-.08em
    T\kern-.1667em\lower.7ex\hbox{E}\kern-.125emX}}

\usepackage[T1]{fontenc}
\usepackage{etoolbox}
\usepackage{comment}
\makeatletter
\def\input@path{{settings}} % isasmathmacros saved in folder './settings/'
\usepackage{settings/isasmathmacros} 
\makeatletter

\usepackage{nicefrac}
\usepackage{array}
\usepackage{booktabs}
\usepackage[linesnumbered,ruled,vlined]{algorithm2e}
\AtBeginEnvironment{algorithm}{\DontPrintSemicolon}

\usepackage[x11names, svgnames, rgb]{xcolor} % load before pgf
\usepackage{graphics} % for pdf, bitmapped graphics files
\usepackage[]{graphicx}  % TODO: Remove draft before final submission
\graphicspath{ {./images/**} } % more paths can be added {./path2/**}

\usepackage[notransparent]{svg}
\usepackage{adjustbox}

\usepackage{tikz}

\usetikzlibrary{patterns, shapes,}
\usepackage{scalerel} % for scaling symbols
\usepackage{relsize} % change text sizes, e.g. in svg 
\usepackage[font=normalsize,skip=1pt]{caption}
\usepackage[skip=1.5pt, font=small]{subcaption} %used with ieeeconf class

\usepackage{balance} % -> \balance at the end of the document in the second column on the last page

\usepackage{stfloats} % bessere Platzierung der Bilder

\usepackage{cite} % multi citatioin \cite{tag3,tag1,tag2} get auto sorted by number
\usepackage{cleveref}

\crefname{figure}{Fig.}{Figs.} % figure prefix 
\crefname{section}{Sec.}{Secs.} % Section prefix 
\crefname{table}{Tab.}{Tabs.} % Tabular prefix 
\crefname{algorithm}{Alg.}{Algs.} % algorithm prefix
\crefname{equation}{}{} % Equation prefix (empty)

\crefname{assumption}{Assumption}{Assumption} % assumption label format
\crefname{myAssumptionCtr}{Assumption}{Assumption} % assumption label format

\crefname{theorem}{Theorem}{Theorems} % theorem label format
\crefname{myTheoremCtr}{Theorem}{Theorems} % theorem label format

\crefname{example}{Example}{Examples} % example label format
\crefname{myExampleCtr}{Example}{Examples} % example label format

\expandafter\def\csname ver@etex.sty\endcsname{3000/12/31}

\usepackage{autonum}

\usepackage{siunitx}
\usepackage{microtype}
\usepackage[all]{nowidow}
\usetikzlibrary{external}
\usepackage{bbold} % \mathbb for lowercase letters and numbers

\newcommand{\ie}{i.e.}
\newcommand{\Ie}{I.e.}
\newcommand{\eg}{e.g.}

\newcommand{\IRp}{\ensuremath{\IR_{+}}}

\newcommand{\indFunc}{\ensuremath{\mathds{1}}}

\definecolor{myblue}{named}{RoyalBlue}
\definecolor{mygreen}{named}{LimeGreen}

\newcommand{\pdf}{\ensuremath{f}} % probability density function

\newcommand{\cemPropPdf}{\ensuremath{\tilde{f}}} % proposal density

\newcommand{\horizon}{\ensuremath{N}_{\mathrm{H}}} % horizon length

\newcommand{\numCEMIter}{\ensuremath{j}_\mathrm{max}} % number of CEM iterations
\newcommand{\numCEMSamples}{\ensuremath{N}_\mathrm{CEM}} % number of CEM samples
\newcommand{\numCEMElite}{\ensuremath{N_\mathrm{e}}} % number of CEM elite samples

\newcommand{\optVarCemScalar}{\ensuremath{\xi}} % scalar optimization variable for CEM
\newcommand{\vVarCem}{\ensuremath{\vec{\optVarCemScalar}}} % decision variable for CEM
\newcommand{\rvVarCem}{\ensuremath{\rvec{\optVarCemScalar}}} % random decision variable for CEM
\newcommand{\dimVarCem}{\ensuremath{d_{{\optVarCemScalar}}}} % dimension of decision variable for CEM
\newcommand{\evVarCem}{\ensuremath{\mean{\vec{\optVarCemScalar}}}} % mean of the proposal distribution
\newcommand{\paramCEM}{\ensuremath{\vtheta}} % parameter vector of proposal density
\newcommand{\lcdVarScalar}{\ensuremath{\xi}} % scalar variable for LCD
\newcommand{\vVarLcd}{\ensuremath{\vec{\lcdVarScalar}}} % variable for LCD
\newcommand{\dimVarLcd}{\ensuremath{d_{{\lcdVarScalar}}}} % dimension of variable for LCD
\usepackage[]{acro}

\acsetup{single=1} % This option determines whether a single appearance of an acronym counts as usage. It might be desirable in such cases that an acronym is simply printed as long form and not added to the list of acronym.
\acsetup{single-style=long} % single-style = long-short|short-long|short|long|footnote. The style of the single appearance of an acronym.

\DeclareAcronym{AL}         {short={AL},	    	long={active learning}}
\DeclareAcronym{ANEES}      {short={ANEES},	    	long={averaged normalized estimation error squared}}
\DeclareAcronym{ANLL}       {short={ANLL},	    	long={average negative log-likelihood}}

\DeclareAcronym{BNN}		{short={BNN},	    	long={Bayesian neural network}}

\DeclareAcronym{CDF}		{short={CDF},	    	long={cumulative density function}}
\DeclareAcronym{CEM}		{short={CEM},	    	long={cross-entropy method}}
\DeclareAcronym{CMA-ES}	    {short={CMA-ES},	    long={covariance matrix adaptation evolution strategy}, long-plural-form={covariance matrix adaptation evolution strategies}}
\DeclareAcronym{CvM}		{short={CvM},	    	long={Cram\'er--von Mises}}

\DeclareAcronym{DDP}		{short={DDP},	    	long={differential dynamic programming}}
\DeclareAcronym{DM}	        {short={DM},	    	long={Dirac mixture}}
\DeclareAcronym{DMD}		{short={DMD},	    	long={Dirac mixture density}}
\DeclareAcronym{dsCEM}      {short={dsCEM},         long={deterministic sampling CEM}}

\DeclareAcronym{ECE}	    	{short={ECE},	    	long={expected calibration error}}
\DeclareAcronym{EKF}		{short={EKF},	    	long={Extended Kalman Filter}}
\DeclareAcronym{EM}	    	{short={EM},	    	long={expectation--maximization}}
\DeclareAcronym{ENCE}	    	{short={ENCE},	    	long={expected normalized calibration error}}
\DeclareAcronym{EP}	    	{short={EP},	    	long={Expectation Propagation}}

\DeclareAcronym{GM}	        {short={GM},	    	long={Gaussian mixture}}
\DeclareAcronym{GEVR}	    {short={GEVR},	    	long={generalized error variance ratio}}
\DeclareAcronym{GUCE}		{short={GUCE},	    	long={generalized \acs{UCE}}}

\DeclareAcronym{iLQR}		{short={iLQR},	    	long={iterative linear quadratic regulator}}
\DeclareAcronym{kdtree}		{short={\mbox{$k$-d~tree}},	    	long={$k$-dimensional tree}}
\DeclareAcronym{KBNN}		{short={KBNN},	    	long={Kalman Bayesian Neural Networks}}
\DeclareAcronym{KL}         {short={KL},            long={Kullback--Leibler}}
\DeclareAcronym{KS}		    {short={KS},	    	long={Kolmogorov--Smirnov}}

\DeclareAcronym{LCD}		{short={LCD},	    	long={localized cumulative distribution}}
\DeclareAcronym{LRKF}	    {short={LRKF},	    	long={Linear Regression Kalman filter}}

\DeclareAcronym{MAP}	    {short={MAP},	    	long={maximum a posteriori}}
\DeclareAcronym{MC}	        {short={MC},	    	long={Monte Carlo}}
\DeclareAcronym{MCMC}	    {short={MCMC},	    	long={Markov Chain
Monte Carlo}}
\DeclareAcronym{ML}		    {short={ML},	    	long={maximum likelihood}}
\DeclareAcronym{MLE}		{short={MLE},	    	long={maximum likelihood estimator}}
\DeclareAcronym{MPC}		{short={MPC},	    	long={model predictive control}}
\DeclareAcronym{MPPI}		{short={MPPI},	    	long={model predictive path integral}}
\DeclareAcronym{MSE}		{short={MSE},	    	long={mean squared error}}
\DeclareAcronym{MMSE}		{short={MMSE},	    	long={minimum mean square error}}

\DeclareAcronym{MNR}		{short={MNR},	    	long={matrix norm ratio}}
\DeclareAcronym{MNRE}		{short={MNRE},	    	long={matrix norm relative  error}}
\DeclareAcronym{MSER}		{short={MSER},	    	long={mean squared error ratio}}

\DeclareAcronym{MV}		{short={MV},	    	long={mean variance}}

\DeclareAcronym{NCI}		{short={NCI},	    	long={noncredibility index}}
\DeclareAcronym{NEES}      {short={NEES},	    	long={normalized estimation error squared}}
\DeclareAcronym{NLL}       {short={NLL},	    	long={negative log-likelihood}}
\DeclareAcronym{NGUCE}      {short={NGUCE},	    	long={normalized \ac{GUCE}}}
\DeclareAcronym{NUTS}      {short={NUTS},	    	long={No-U-Turn Sampler}}

\DeclareAcronym{ODE}	    {short={ODE},	    	long={ordinary differential equation}}
\DeclareAcronym{OCP}	    {short={OCP},	    	long={optimal control problem}}

\DeclareAcronym{PBP}		{short={PBP},	    	long={probabilistic backpropagation}}

\DeclareAcronym{PCD}		{short={PCD},	    	long={projected cumulative distribution}}
\DeclareAcronym{PDF}		{short={PDF},	    	long={probability density function}}

\DeclareAcronym{QCE}		{short={QCE},	    	long={quantile calibration error}}

\DeclareAcronym{RMSE}		{short={RMSE},	    	long={root \ac{MSE}}}
\DeclareAcronym{RMV}		{short={RMV},	    	long={root \ac{MV}}}
\DeclareAcronym{RNN}		{short={RNN},	    	long={recurrent neural network}}
\DeclareAcronym{S2KF}	    {short={S$^2$KF},	    long={Smart Sampling Kalman Filter}}
\DeclareAcronym{SVI}		{short={SVI},	    	long={Stochastic Variational Inference}}
\DeclareAcronym{TAGI}		{short={TAGI},	    	long={Tractable Approximate Gaussian Inference}}

\DeclareAcronym{UCE}		{short={UCE},	    	long={uncertainty calibration error}}
\DeclareAcronym{UKF}		{short={UKF},	    	long={Unscented Kalman Filter}}

\DeclareAcronym{VI}		    {short={VI},	    	long={Variational Inference}}

\title{%
    \LARGE\bf% ieeeconf
    Sample-Efficient and Smooth Cross-Entropy Method\\%
    Model Predictive Control Using Deterministic Samples%
\author{Markus Walker, Daniel Frisch, and Uwe D. Hanebeck}
\thanks{%
    This work is part of the German Research Foundation (DFG) AI Research Unit 5339 regarding the combination of physics-based simulation with AI-based methodologies for the fast maturation of manufacturing processes.}% <-this % stops a space
\thanks{%
    Markus~Walker, Daniel~Frisch and Uwe~D.~Hanebeck are with the Intelligent Sensor-Actuator-Systems Laboratory (ISAS), Institute for Anthropomatics and Robotics, Karlsruhe Institute of Technology, Germany (e-mail:
    \{%
        {\tt\footnotesize markus.walker},
        {\tt\footnotesize daniel.frisch}, 
        {\tt\footnotesize uwe.hanebeck}%
    \}%
    {\tt\footnotesize @kit.edu}%
    ).%
    }% <-this % stops a space
}

\begin{document}

\maketitle
\thispagestyle{empty}
\pagestyle{empty}

%%%%%%%%%%%%%%%%%%%%%%%%%%%%%%%%%%%%%%%%%%%%%%%%%%%%%%%%%%%%%%%%%%%%%%%%%%%%%%%%
% abstract
%%%%%%%%%%%%%%%%%%%%%%%%%%%%%%%%%%%%%%%%%%%%%%%%%%%%%%%%%%%%%%%%%%%%%%%%%%%%%%%%

\begin{abstract}
    Cross-entropy method model predictive control (CEM--MPC) is a powerful gradient-free technique for nonlinear optimal control, but its performance is often limited by the reliance on random sampling.
    This conventional approach can lead to inefficient exploration of the solution space and non-smooth control inputs, requiring a large number of samples to achieve satisfactory results.
    To address these limitations, we propose \ac{dsCEM}, a novel framework that replaces the random sampling step with deterministic samples derived from \acp{LCD}.
    Our approach introduces modular schemes to generate and adapt these sample sets, incorporating temporal correlations to ensure smooth control trajectories.
    This method can be used as a drop-in replacement for the sampling step in existing \acs{CEM}-based controllers.
    Experimental evaluations on two nonlinear control tasks demonstrate that \ac{dsCEM} consistently outperforms state-of-the-art iCEM in terms of cumulative cost and control input smoothness, particularly in the critical low-sample regime.
    %
    % reset acronyms after abstract
    \acresetall
\end{abstract}

%%%%%%%%%%%%%%%%%%%%%%%%%%%%%%%%%%%%%%%%%%%%%%%%%%%%%%%%%%%%%%%%%%%%%%%%%%%%%%%%

%%%%%%%%%%%%%%%%%%%%%%%%%%%%%%%%%%%%%%%%%%%%%%%%%%%%%%%%%%%%%%%%%%%%%%%%%%%%%%%%
% key words
%%%%%%%%%%%%%%%%%%%%%%%%%%%%%%%%%%%%%%%%%%%%%%%%%%%%%%%%%%%%%%%%%%%%%%%%%%%%%%%%
% \begin{IEEEkeywords}
\begin{keywords}
    Model predictive control, cross-entropy method, deterministic sampling, localized cumulative distribution, sampling efficiency.
\end{keywords}
% \end{IEEEkeywords}
%%%%%%%%%%%%%%%%%%%%%%%%%%%%%%%%%%%%%%%%%%%%%%%%%%%%%%%%%%%%%%%%%%%%%%%%%%%%%%%%

% \todo[inline]{Dieser Kommentar wird als farbig hinterlegter Absatz in den Textfluss eingefügt} 
% \todo{Dieser Kommentar wird auf den Seitenrand gequetscht.}  

%%%%%%%%%%%%%%%%%%%%%%%%%%%%%%%%%%%%%%%%%%%%%%%%%%%%%%%%%%%%%%%%%%%%%%%%%%%%%%%%
% Sections
%%%%%%%%%%%%%%%%%%%%%%%%%%%%%%%%%%%%%%%%%%%%%%%%%%%%%%%%%%%%%%%%%%%%%%%%%%%%%%%%

\section{INTRODUCTION}
\label{sec:intro}

Optimal control plays a crucial role in achieving desired performance in many applications, such as autonomous driving and robotics.
One widely used method for solving these types of problems is \ac{MPC}, which operates in a receding horizon fashion.
At each time step, \ac{MPC} solves an optimization problem to determine the optimal control inputs over a finite horizon. It then applies only the first input to the system~\cite{rawlingsModel2017}.
However, this optimization can be challenging for nonlinear systems, non-differentiable dynamics, and non-convex cost functions.
In such cases, gradient-free optimization methods are often preferred because they can navigate complex cost landscapes and avoid local minima~\cite{pinneriSampleefficientCrossentropyMethod2021}.

\begin{figure}
    \centering
    \includegraphics[width=0.95\columnwidth]{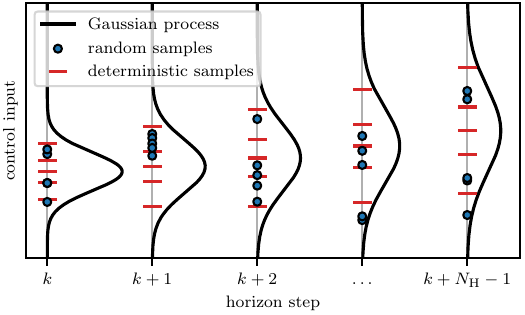} 
    \caption{%
    Schematic showing control input sampling over a finite horizon using either deterministic or random samples. As can be seen, deterministic samples cover the stochastic process without large gaps or clusters. 
    For simplicity, time correlations are neglected.
    }%
    \label{fig:eye_catcher}
\end{figure}

The \ac{CEM}~\cite{rubinsteinCrossentropyMethodCombinatorial1999} is a popular gradient-free technique that has been successfully applied to various optimal control problems~\cite{pinneriSampleefficientCrossentropyMethod2021,chuaDeepReinforcementLearning2018}.
It iteratively refines a proposal distribution over control inputs by sampling, evaluating performance, and updating the distribution based on the best-performing samples.
\ac{CEM}--\ac{MPC} integrates this into the \ac{MPC} framework, using a discrete-time stochastic process (\eg, a Gaussian process, see \cref{fig:eye_catcher}) to generate control input sequences.
The parameters of this process are updated based on the cost of each sampled trajectory, which focuses the search on promising regions of the control space.

A key challenge in standard \ac{CEM}--\ac{MPC} is that control inputs are sampled independently at each time step, which can result in non-smooth control sequences and requires a large number of samples to adequately explore the control space.
The improved \ac{CEM} (iCEM)~\cite{pinneriSampleefficientCrossentropyMethod2021} introduces smoothness by temporal correlations, allowing for more informed sampling strategies that lead to better performance.
However, reliance on random sampling, even with temporal correlations, can still produce non-smooth control trajectories.
This is often undesirable in practice because it can damage actuators and cause erratic system behavior.

To overcome these challenges, we propose replacing random sampling with a deterministic sampling strategy.
Our approach, which we term \emph{deterministic sampling Cross-Entropy Method} (\acs{dsCEM}), leverages precomputed optimal sample sets based on \acp{LCD}~\cite{MFI08_Hanebeck-LCD,CDC09_HanebeckHuber}.
These deterministic samples are structured to cover the solution space more efficiently and with fewer discrepancies than random samples (see~\cref{fig:eye_catcher}).
Inspired by iCEM, we incorporate temporal correlations into the sampling process with the goal of generating control sequences that are both more effective and significantly smoother.
The proposed sampling schemes are designed as modular drop-in replacements for the sampling step in \ac{CEM}-based controllers, making this approach broadly applicable.
\subsection{Contribution And Outline}
We first state the optimal control problem and summarize the \ac{CEM}--\ac{MPC} approach in \cref{sec:ocp,sec:cem_mpc}.
Then, in \cref{sec:lcd_samples}, we introduce the concept of deterministic samples based on \acp{LCD} and describe how to integrate them into the \ac{CEM}--\ac{MPC} framework.
Finally, we evaluate the proposed approach on two benchmark tasks, (i)~the mountain car and (ii)~the cart-pole swing-up task, in \cref{sec:evaluation}, and compare it against the widely used iCEM method~\cite{pinneriSampleefficientCrossentropyMethod2021}.

\subsection{Notation}
In this paper, underlined letters, \eg, $\vx$, denote vectors, boldface letters, such as $\rvx$, represent random variables, while boldface capital letters, such as $\mA$, indicate matrices.
Sets are denoted by calligraphic letters, \eg, $\mathcal{E}$.
The mean of a random variable is denoted by $\hat{\cdot}$, \eg, $\hat{\rvx}$, and covariance matrices are denoted by $\mC$.
Diagonal matrices are represented by $\diag(\cdot)$.
The indicator function is denoted by $\indFunc$, \eg, $\indFunc_{A}$ is \num{1} if event $A$ is true and \num{0} otherwise.
% Expectation operators are denoted by $\E{\cdot}$; \eg, $\E[p(\vx)]{\rvx}$ denotes the expectation of $\rvx$ under the \ac{PDF} $p(\vx)$. When appropriate, the density argument is omitted for simplicity.
% The same convention applies to the variance operator, denoted by $\Var{\cdot}$.
\section{OPTIMAL CONTROL PROBLEM}
\label{sec:ocp}

We consider discrete-time, finite-horizon deterministic \acp{OCP} with cumulative cost
\begin{align}
    \label{eq:ocp_cost}
    J_k & = g_{\horizon}(\vx_{k + \horizon}) + \sum_{n=k}^{k+\horizon-1} g_n(\vx_n, \vu_n)  \enspace,
\end{align}
where $k$ is the current time step, $\horizon$ the prediction horizon, $\vx_k \in \IR^{d_x}$ the system state, $\vu_n \in \IR^{d_u}$ the control input, $g_n(\vx_n, \vu_n)\colon \IR^{d_x} \times \IR^{d_u} \to \IR$ the stage cost at time step $n$, and $g_{\horizon}(x_{k+\horizon})\colon \IR^{d_x} \to \IR$ the terminal cost. 
The discrete-time system dynamics are given by
\begin{align}
    \label{eq:system_dynamics}
    \vx_{n+1} & = \va_n(\vx_n, \vu_n) \quad \text{for } n=k,\ldots,k+\horizon-1 \enspace, 
\end{align}
with initial state $\vx_k$ at time step $k$ and system function $\va_n(\vx_n, \vu_n)\colon \IR^{d_x} \times \IR^{d_u} \to \IR^{d_x}$. 
The \ac{OCP} is to find the optimal control input sequence $\vu_k^*, \ldots, \vu_{k+\horizon-1}^*$ that minimizes the cumulative cost $J_k$ while satisfying system dynamics and additional constraints on states and controls. 
The \ac{OCP} can be solved in a receding horizon fashion using \ac{MPC}, where at each time step $k$, the first control input $\vu_k^*$ of the optimal sequence is applied to the system, and the process is repeated at the next time step.

Typically, the \ac{OCP} is solved using gradient-based approaches such as direct single- or multiple-shooting methods~\cite{diehlFast2006}, \ac{DDP}~\cite{jacobsonDifferentialDynamicProgramming1970}, \ac{iLQR}~\cite{liIterativeLinearQuadratic2004}, or gradient-free methods such as \acp{MPPI}~\cite{williamsAggressiveDrivingModel2016}, 
% \acp{CMA-ES}~\cite{Hansen2006}, 
or the \ac{CEM}~\cite{chuaDeepReinforcementLearning2018,pinneriSampleefficientCrossentropyMethod2021}. 
In this work, we focus on \ac{CEM}--\ac{MPC}, which is described in more detail below.
\section{CROSS-ENTROPY METHOD MPC}
\label{sec:cem_mpc}

Before introducing the \ac{CEM}--\ac{MPC}, we will briefly explain the concept of the \ac{CEM} as an optimizer, based on~\cite{rubinsteinCrossentropyMethodCombinatorial1999, deboerTutorialCrossentropyMethod2005}.
Note that, to prevent confusion with the control input $\vu$ and state $\vx$, we denote the $\dimVarCem$-dimensional optimization variable for the \ac{CEM} as $\vVarCem \in \IR^{\dimVarCem}$.

\subsection{Cross-Entropy Method}
\label{sec:cem}

% The background of the \ac{CEM} is summarized below, based on~\cite{rubinsteinCrossentropyMethodCombinatorial1999, deboerTutorial2005}.

The \ac{CEM} aims to find a density $\rvVarCem\sim\pdf(\vVarCem; \paramCEM)$, parameterized by $\paramCEM$, whose samples $\vVarCem^{(i)}$ tend to achieve small costs $J(\vVarCem^{(i)})$. 
% The \ac{CEM} can be viewed as the estimation of rare-event probabilities.
% The rare events correspond to finding solutions $\vVarCem^{(i)}$ that are are realizations of the random vector $\rvVarCem\sim\pdf(\vVarCem; \paramCEM)$, which is parameterized by $\paramCEM$,
% and that achieve costs $J(\vVarCem^{(i)})$ of an unconstrained minimization problem below a certain threshold.
%  achieve low costs $J(\vu_i)$ of an unconstrained minimization problem below a certain threshold, where $\vu_i$ are realizations of the random vector $\rvu\sim\pdf(\vu; \vtheta)$, which is parameterized by $\vtheta$.
By setting a performance threshold $\gamma$ and defining the event $A_\gamma = \{ \vVarCem \mid J(\vVarCem) \le \gamma \}$, \ie, the event of achieving low-cost solutions, the corresponding rare-event probability can be expressed as~\cite{rubinsteinCrossentropyMethodCombinatorial1999}
\begin{align}
    l = \Pr_{\pdf(\vVarCem; \paramCEM)}\!\big( J(\rvVarCem) \le \gamma \big) 
    = \Eop_{\pdf(\vVarCem; \paramCEM)} \big\{ \indFunc_{J(\rvVarCem)\le \gamma} \big \} \enspace,
    \label{eq:rare_event_prob}
\end{align}
where $\indFunc_{J(\vVarCem)\le \gamma}$ is the indicator function.
%  that is $1$ if {$J(\vu) \le \gamma$} and $0$ otherwise.
By multiplying $\nicefrac{\cemPropPdf(\vVarCem)}{\cemPropPdf(\vVarCem)}$ within the expectation \cref{eq:rare_event_prob}, rearranging yields
\begin{align}
    l = \Eop_{\pdf(\vVarCem; \paramCEM)} \left\{ \indFunc_{J(\rvVarCem)\le \gamma} \frac{\cemPropPdf(\vVarCem)}{\cemPropPdf(\vVarCem)} \right\} 
    = \Eop_{\cemPropPdf(\vVarCem)} \left\{ \indFunc_{J(\rvVarCem)\le \gamma} \frac{\pdf(\vVarCem; \paramCEM)}{\cemPropPdf(\vVarCem)} \right\} 
    \enspace,
    % \frac{\cemPropPdf(\vu;\vtheta)}{\pdf(\vu)} \right] \enspace.
\end{align} 
where $\cemPropPdf(\vVarCem)$ is an importance sampling \ac{PDF}. 
Note that the expectation is now w.r.t. the importance \ac{PDF} $\cemPropPdf(\vVarCem)$.
According to importance sampling theory, the optimal choice of $\cemPropPdf^\ast(\vVarCem)$ that minimizes the variance of the estimator is~\cite{rubinsteinCrossentropyMethodCombinatorial1999}
\begin{align}
    \cemPropPdf^*(\vVarCem) = \frac{\indFunc_{J(\vVarCem)\le \gamma} \, \pdf(\vVarCem;\paramCEM)}{l} \enspace.
\end{align}
However, this is infeasible because it depends on $l$, which is the quantity we want to estimate using $\cemPropPdf(\vVarCem)$, and is therefore a-priori unknown.

To overcome this infeasibility, the central idea of \ac{CEM} is to \emph{iteratively approximate} the optimal \ac{PDF} $\cemPropPdf^*(\vVarCem)$. 
At each iteration $j$, we start with a known proposal distribution $\cemPropPdf(\vVarCem;\paramCEM_j)$ and aim to find a better set of parameters $\paramCEM_{j+1}$ for the next iteration.
This is achieved by finding the parameters $\paramCEM_{j+1}$ that minimize the \ac{KL} divergence between the 
% (hypothetical) 
optimal \ac{PDF} derived from the current step, 
\begin{align}
\cemPropPdf_j^*(\vVarCem) = \frac{\indFunc_{J(\vVarCem)\le \gamma_j} \, \cemPropPdf(\vVarCem;\paramCEM_j)}{l_j} \enspace,
\end{align}
and the next proposal distribution $\cemPropPdf(\vVarCem;\paramCEM_{j+1})$. 
Formally, the next parameters are given by
\begin{align}
        \paramCEM_{j+1} &= \argmin_{\vtheta'} D_{\mathrm{KL}}\mleft(\cemPropPdf_j^*(\vVarCem) \| \cemPropPdf(\vVarCem;\vtheta') \mright) \\
        & = \argmin_{\vtheta'} H\mleft(\cemPropPdf_j^*(\vVarCem), \cemPropPdf(\vVarCem;\vtheta') \mright) - H\mleft(\cemPropPdf_j^*(\vVarCem)\mright) 
        % \\
        % & = \int \pdf(u) \log \left( \frac{\pdf(u)}{\cemPropPdf(u; \vtheta)} \right) \dd \vu
\end{align}
where $H(\cdot,\cdot)$ is the cross-entropy between both distributions, and $H(\cemPropPdf_j^*(\vVarCem))$ is the entropy of $\cemPropPdf_j^*(\vVarCem)$.
Since entropy $H(\cdot)$ does not depend on $\vtheta'$, minimizing the \ac{KL} divergence is equivalent to minimizing the cross-entropy
\begin{align}
    \paramCEM_{j+1} &= \argmin_{\vtheta'} - \int_{\IR^{\dimVarCem}} \cemPropPdf_j^\ast(\vVarCem) \log \cemPropPdf(\vVarCem; \vtheta') \dd \vVarCem \\
    &= \argmin_{\vtheta'} - \int_{\IR^{\dimVarCem}} \frac{\indFunc_{J(\vVarCem)\le \gamma_j} \, \cemPropPdf(\vVarCem;\paramCEM_j)}{l_j} \log \cemPropPdf(\vVarCem; \vtheta') \dd \vVarCem \enspace,
\end{align}
where the unknown constant $l_j$ can be dropped, as it does not depend on $\vtheta'$. 
By rewriting the integral as an expectation with respect to the \emph{known} current \ac{PDF} $\cemPropPdf(\vVarCem;\paramCEM_j)$, we get
\begin{align}
    \paramCEM_{j+1} = \argmin_{\vtheta'} - \Eop_{\cemPropPdf(\vVarCem; \paramCEM_j)} \big\{ \indFunc_{J(\rvVarCem)\le \gamma_j} \log \cemPropPdf(\vVarCem; \vtheta') \big\} \enspace.
    \label{eq:cem_update_expectation}
\end{align}
This expectation can be approximated by samples $\{\vVarCem^{(i)}\}_{i=1}^N$ from $\cemPropPdf(\vVarCem;\paramCEM_j)$. 
This yields the objective for the next set of parameters
\begin{align}
    \paramCEM_{j+1} = \argmin_{\vtheta'} - \frac{1}{N} \sum_{i=1}^N \indFunc_{J(\vVarCem^{(i)})\le \gamma_j} \log \cemPropPdf(\vVarCem; \vtheta') \enspace.
\end{align}
Therefore, the new parameters $\vtheta'$ are fitted to the subset of best samples 
$\mathcal{E}_j = \{ \vVarCem^{(i)} \mid J(\vVarCem^{(i)}) \le \gamma_j \}$,
% for which satisfy $J(\vu_i) \le \gamma_j$, 
\ie, the \emph{elite set}.

For common choices of $\cemPropPdf$, the minimization has simple closed-form solutions, \eg, for a Gaussian $\cemPropPdf(\vVarCem;\paramCEM) = \mathcal{N}(\vVarCem ; \evVarCem, \mC)$, where the solution is the maximum likelihood estimate on the elite set
\begin{align}
    \label{eq:cem_mean_update}
    \evVarCem_{j+1} & = \frac{1}{|\mathcal{E}_j|} \sum_{\vVarCem^{(i)} \in \mathcal{E}_j} \vVarCem^{(i)} 
    \enspace,
    \\
    \label{eq:cem_cov_update}
    \mC_{j+1} & = \frac{1}{|\mathcal{E}_j|} \sum _{\vVarCem^{(i)} \in \mathcal{E}_j} (\vVarCem^{(i)} - \evVarCem_{j+1}) (\vVarCem^{(i)} - \evVarCem_{j+1})\T \enspace,
\end{align}
with $\evVarCem_{j+1}$ and $\mC_{j+1}$ being the mean and covariance matrix of the Gaussian proposal summerized in parameter vector $\paramCEM_{j+1} = (\evVarCem_{j+1}, \mC_{j+1})$ and the elite set size $|\mathcal{E}_j|$.

In each iteration, $\gamma_j$ is reduced to focus on better-performing samples (\ie, more rare events).
In practice, $\gamma_j$ is usually not specified directly but rather, the $\numCEMElite$-th best cost among the samples is used, \eg, the best $\num{10}$ samples out of $\num{100}$.
After the last iteration $\numCEMIter$, the best sample from the final elite set $\mathcal{E}_{\numCEMIter}$ is returned as the solution.
Note that in some variants, the mean $\evVarCem_{\numCEMIter}$ of the final proposal distribution is returned instead.

In summary, the \ac{CEM} is an iterative method that refines the proposal distribution over the solution space, concentrating on regions with lower costs, and effectively guiding the search toward optimal or near-optimal solutions. 
The complete \ac{CEM} procedure is shown in \cref{alg:cem}.

% Short summary: 

% The \ac{CEM} \cite{rubinsteinCrossEntropyMethod1999} is a optimization method that iteratively refines a proposal distribution $\cemPropPdf(\vu; \vtheta_k)$ over the solution space to concentrate on regions with lower costs.

\begin{algorithm}[t]
    \caption{Cross-Entropy Method}
    \label{alg:cem}
    % define as function
    \SetKwFunction{FMain}{CEM}

    \SetKwInOut{Input}{Input}
    \Input{Initial parameters $\paramCEM_0$, elite set size $\numCEMElite$, number of samples $\numCEMSamples$, number of iterations $\numCEMIter$}
    \SetKwInOut{Output}{Output}
    \For{$j \gets 0$ \KwTo $\numCEMIter-1$}{
        Sample $\{\vVarCem^{(i)}\}_{i=1}^{\numCEMSamples} \sim \cemPropPdf(\vVarCem; \paramCEM_j)$\;
        Evaluate costs $\{J(\vVarCem^{(i)})\}_{i=1}^{\numCEMSamples}$\;
        % Determine threshold $\gamma_j$ from the $\numCEMElite$-th best cost\;
        % Select elite set $\mathcal{E}_j = \{ \vu_i \mid J(\vu_i) \le \gamma_j \}$\;
        Select elite set $\mathcal{E}_j$ of the $\numCEMElite$ best samples\;
        Update parameters $\paramCEM_{j+1}$ using $\mathcal{E}_j$\;
    }
    \Return best $\vVarCem^{(i)}$ from final elite set $\mathcal{E}_{\numCEMIter}$

\end{algorithm}

\subsection{Application to MPC}

To apply the \ac{CEM} to the \ac{OCP} from~\cref{sec:ocp}, the entire control sequence over the prediction horizon $\rvu_{k:k+\horizon-1} = \rvu_k, \ldots, \rvu_{k+\horizon-1}$, is treated (in flattened form) as the random vector to be optimized at each time step $k$. 
% \Ie, the optimization variable is $\rvVarCem = [\rvu_k\T, \ldots, \rvu_{k+\horizon-1}\T]$ over $\IR^{d_u \cdot H-1}$, and $J_k$~\cref{eq:ocp_cost} is the cost function which is to be minimized.
\Ie, the optimization variable is the random vector $\rvVarCem = [\rvu_k\T, \ldots, \rvu_{k+\horizon-1}\T]\T$, whose sample space is $\IR^{d_u \cdot \horizon}$, and the goal is to find a realization that minimizes the cost function $J_k$~\cref{eq:ocp_cost}.

The sequence of control inputs can be viewed as a discrete-time stochastic process. 
A natural choice for the proposal distribution $\cemPropPdf(\cdot)$ is therefore a multivariate Gaussian distribution over the flattened control sequence
% \begin{align}
    $\cemPropPdf(\vVarCem; \vtheta_j) = \mathcal{N}(\vVarCem ; \evVarCem_j, \mC_j)$,
% \end{align}
where the parameters $\vtheta_j = (\evVarCem_j, \mC_j)$ consist of the mean control sequence $\evVarCem_j = [\evu_k\T, \ldots, \evu_{k+\horizon-1}\T]\T$  and the covariance matrix $\mC_j \in \IR^{(d_u \cdot \horizon) \times (d_u \cdot \horizon)}$. 
This formulation is powerful because the covariance matrix $\mC_j$ can model temporal correlations between control inputs at different time steps, effectively treating the control sequence as a Gaussian Process.
As with standard \ac{CEM}, the parameters are iteratively updated based on the elite set of best-performing control sequences. 
Notably, trajectory shooting and cost evaluation can be \emph{efficiently parallelized} across all sampled control sequences on modern hardware.
The overall procedure for a single \ac{CEM}--\ac{MPC} step is summarized in \cref{alg:cem_mpc}.

\begin{algorithm}[t]
    \caption{Cross-Entropy Method MPC Step}
    \label{alg:cem_mpc} 
    \SetKwFunction{FMain}{CEM-MPC-Step}
    \SetKwInOut{Input}{Input}
    \Input{%
        State $\vx_k$, initial parameters $\paramCEM_0 = (\evVarCem_0, \mC_0)$
    }%
    \SetKwInOut{Output}{Optimal control input $\vu_k^*$}
    \For{$j \gets 0$ \KwTo $\numCEMIter-1$}{
        % Sample $\{\vVarCem^{(i)}\}_{i=1}^{\numCEMSamples} \sim \cemPropPdf(\cdot; \paramCEM_j)$\;
        Sample $\vu^{(1)}_{k:k+\horizon-1}, \ldots, \vu^{(\numCEMSamples)}_{k:k+\horizon-1} \sim \cemPropPdf(\cdot; \paramCEM_j)$\; \label{alg:cem_mpc:sampling_step}
        Trajectory shooting using \cref{eq:system_dynamics}~\tcp{\small{parallel}}
        Evaluate costs $\{J_k(\vu^{(i)}_{k:k+\horizon-1})\}_{i=1}^{\numCEMSamples}$ using~\cref{eq:ocp_cost}~\tcp{\small{parallel}}
        Select elite set $\mathcal{E}_j$ of the $\numCEMElite$ best samples\;
        Update parameters $\paramCEM_{j+1}$ using $\mathcal{E}_j$ \;
    }
    \Return first control $\vu_k^*$ from the best sequence in $\mathcal{E}_{\numCEMIter}$
\end{algorithm}

\subsection{Practical Improvements and Related Work}
\label{sec:cem_mpc:related_work}

A major challenge in this basic approach is the high dimensionality of the optimization problem. 
While adapting a full covariance matrix can capture correlations, it is prone to estimation errors unless a large number of samples is used.
To address this, a common simplification is to assume a diagonal covariance matrix $\mC_j$, \eg, as done in~\cite{hafnerLearning2019}, which reduces the number of parameters to be estimated and simplifies the sampling process, \ie, $\mC_j = \diag(\sigma_{j,1}^2, \ldots, \sigma_{j,d_u \cdot \horizon}^2)$, where $\sigma_{j,i}^2$ is the variance of the $i$-th control input in the flattened sequence.
% If the covariance matrix $\mC_j$ is diagonal, the control inputs at each time step are sampled independently (white noise), which often leads to noisy control actions. 
However, this assumption ignores temporal correlations between control inputs at different time steps, which can be crucial for generating smooth control sequences.

A standard technique to improve convergence smoothness is to introduce a momentum term in the parameter updates~\cite{deboerTutorialCrossentropyMethod2005}, which smooths the updates across iterations. The momentum update is given by
% \begin{align}
%     \label{eq:cem_momentum}
    $\evVarCem_{j+1} = \alpha \evVarCem_{j} + (1-\alpha) \evVarCem_{j,\mathrm{e}}$, %\enspace,
% \end{align}
where $\evVarCem_{j}$ is the mean from the previous iteration, $\evVarCem_{j,\mathrm{e}}$ is the mean of the current elite set, and $\alpha \in [0,1)$ is a momentum factor. 
Analogously, the momentum can also be applied to the covariance matrix update~\cite{deboerTutorialCrossentropyMethod2005}.

A further standard improvement, used in various \ac{MPC} approaches~\cite{rawlingsModel2017}, is to warm-start the optimization at each time step by shifting the mean control sequence from the previous time step. This is achieved by setting the initial mean to $\evVarCem_0 = [\vu_{k-1,1}\T, \ldots, \vu_{k-1,\horizon-1}\T, \vu_{\mathrm{init}}\T]\T$, where $\vu_{k-1,i}$ is the $i$-th control input from the optimized sequence at the previous time step and $\vu_{\mathrm{init}}$ is an initial guess for the last control input, often set to zero or the last optimized input.

A notable improvement is the iCEM algorithm~\cite{pinneriSampleefficientCrossentropyMethod2021}, which introduces temporal correlations in a structured way. 
Instead of adopting a full covariance matrix, iCEM samples control actions from colored noise, imposing a smoothness prior on the control sequences. 
This is achieved by generating samples from a stationary distribution whose power spectral density follows a power law, $PSD(f) \propto \nicefrac{1}{f^\beta}$, where $f$ is frequency and $\beta$ controls the noise color.
To further increase sample efficiency, iCEM retains a fraction of the elite set from the previous \ac{MPC} time step, shifting the sequences in time and adding them to the current sample pool.
Other improvements in iCEM include clipping control samples to ensure feasibility and decaying the number of samples over iterations via $N_{\mathrm{CEM}, j} = \max(\nicefrac{\numCEMSamples}{\eta^{j}}, 2\numCEMElite)$, where $\eta \ge 1$ is a decay factor. 
This reduces computational cost as the optimizer converges.
An overview of the iCEM algorithm is provided in~\cite[Alg. 1]{pinneriSampleefficientCrossentropyMethod2021}.

Besides the improvements mentioned above, further enhancements focus on learning the sampling distribution itself, for instance, by using Gaussian processes~\cite{watsonInferringSmoothControl2023} or normalizing flows~\cite{powerLearningGeneralizableTrajectory2024}.

However, in this work, we focus on the sample generation process itself by introducing deterministic sampling to \ac{CEM}--\ac{MPC}. This approach is orthogonal to the learning-based improvements and can be combined with them.
\section{Cross-Entropy Method MPC Using Deterministic Samples}
\label{sec:lcd_samples}

As an alternative to random sampling, we propose to use deterministic samples based on the \ac{LCD}. In this section, we briefly summarize how to obtain such optimal samples and then describe how to integrate them into \ac{CEM}--\ac{MPC}.

% As an alternative, in [8], the use
% of deterministic samples based on the Localized Cumulative
% Distribution (LCD) is proposed. 

% Deterministic LCD samples
% exhibit superior space coverage and homogeneity when
% compared to random samples.

\subsection{Deterministic Samples Using Localized Cumulative Distributions}
\begin{figure}[t]
    \centering
    \begin{subfigure}{0.49\columnwidth}
        \includegraphics{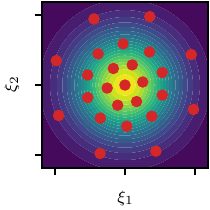}%
        \caption{Standard normal samples}%
        \label{fig:lcd_samples:standard}%
    \end{subfigure}
    \begin{subfigure}{0.49\columnwidth}
        \includegraphics{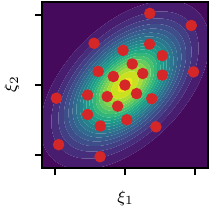}%
        \caption{Transformed samples}%
        \label{fig:lcd_samples:transformed}%
    \end{subfigure}%
    \caption{Example of \num{25} two-dimensional deterministic samples, where the background color indicates the \ac{PDF}.}%
    \label{fig:lcd_samples}%
\end{figure}

The \ac{LCD} is a multivariate generalization \cite{MFI08_Hanebeck-LCD} of the univariate
\ac{CDF}. It is defined as
\begin{align}
    F(\vm, b) & = \int_{\IR^{\dimVarLcd}} f(\vVarLcd) K(\vVarLcd, \vm, b) \dd\vVarLcd \enspace,
\end{align}
where $f(\vVarLcd)$ is a \ac{PDF} over $\IR^{\dimVarLcd}$,  and $K(\cdot, \vm, b)$ is a kernel function centered at $\vm \in \IR^{\dimVarLcd}$ with bandwidth $b \in \IRp$. The kernel function is typically chosen as a Gaussian kernel \cite{MFI08_Hanebeck-LCD} 
\begin{align}
    K(\vVarLcd, \vm, b) & = \exp\left(-\frac{\|\vVarLcd - \vm\|_2^2}{2b^2}\right) \enspace.
\end{align}
We then obtain a general distance measure between two \acp{PDF} by comparing their respective \acp{LCD} with a modified \ac{CvM} distance  \cite{MFI08_Hanebeck-LCD}
\begin{align}
    D_{\operatorfont CvM} & = \int_{0}^{\infty} w(b) \int_{\IR^{\dimVarLcd}} \left( \tilde{F}(\vm, b) - F(\vm, b) \right)^2 \dd \vm \dd b \enspace,
\end{align}
where $\tilde{F}(\vm, b)$ and $F(\vm, b)$ are the \acp{LCD} of $\tilde{f}(\vVarLcd)$ and $f(\vVarLcd)$, respectively, and $w(b)$ is a weighting function, typically $w(b)=b^{1-\dimVarLcd}$. 
Here we use this distance to compare a Gaussian \ac{PDF} $\tilde f(\vVarLcd)$ with a Dirac mixture ${f}(\vVarLcd) =  \frac{1}{N} \sum_{i=1}^N\delta(\vVarLcd - \vVarLcd^{(i)})$. By minimizing $D_{\operatorfont CvM}$ w.r.t. the sample locations $\vVarLcd^{(i)}$ we obtain an optimal Dirac mixture approximation of the Gaussian $\tilde f(\vVarLcd)$ \cite{CDC09_HanebeckHuber}. 
An example of such optimal samples in two dimensions is shown in~\cref{fig:lcd_samples}.

\subsection{Integration of Deterministic Samples into CEM--MPC}
\label{sec:cem_mpc_lcd_integration}

The core idea is to use deterministic samples rather than random ones in the \ac{CEM}.
To speed up, we use pre-computed deterministic samples since online computation is prohibitive.
Optimal samples $\{\tilde{\vVarLcd}^{(i)}\}$ are generated offline for the isotropic standard Gaussian $\mathcal{N}(\vVarCem;\vzero, \mI)$~\cite{Fusion13_Steinbring, JAIF14_Steinbring-S2KF}. 
These samples are then transformed at runtime (\cref{alg:cem_mpc}, \cref{alg:cem_mpc:sampling_step}) to match the current proposal distribution $\Gaussian(\vVarCem;\evVarCem_j, \mC_j)$ of the \ac{CEM} optimizer via~\cite{JAIF14_Steinbring-S2KF}
\begin{align}
    \label{eq:lcd_sample_transform}
    \vVarCem^{(i)} = \evVarCem_j + \mL_j \tilde{\vVarLcd}^{(i)} \enspace,
\end{align}
where $\mL_j$ is the matrix square root of the Gaussian process covariance matrix, \ie, $\mC_j = \mL_j \mL_j\T$. 
While this transformation does not preserve the samples' optimality in the \ac{CvM} sense~\cite{JAIF14_Steinbring-S2KF} for a non-isotropic target distribution, it provides a practical and efficient way to generate structured samples with low discrepancy.

A naive implementation would use the same transformed set of deterministic samples in every iteration of the optimizer. 
However, we found this limits exploration and can lead to premature convergence, especially with a small number of samples. 
To address this, we propose three schemes for generating varied sample sets across iterations.

\paragraph{Sample Set Variability Schemes}
To enhance exploration, we introduce variability into the sampling process either across iterations or across time steps.

(V1) \textbf{Random Rotation}: 
In each iteration $j$ of the \ac{CEM}, the pre-computed isotropic standard Gaussian samples $\{\tilde{\vVarLcd}^{(i)}\}$ are rotated by a random rotation matrix $\mR_j$ 
% \in SO(\dimVarCem)$ 
before the transformation.
The matrix is drawn from the special orthogonal group $SO(\dimVarCem)$, which consists of all $\dimVarCem \times \dimVarCem$ orthogonal matrices with a determinant of $+1$~\cite{leonStatisticalModelRandom2006}.
Therefore the transformation in \cref{eq:lcd_sample_transform} changes to $\vVarCem^{(i)} = \evVarCem_j + \mL_j \mR_j \tilde{\vVarCem}^{(i)}$.
Note that rotating the samples of an isotropic distribution results in another valid sample set without losing their optimality in the \ac{CvM} sense. % 
This introduces stochasticity, improving exploration at the cost of losing the fully deterministic nature of the algorithm.

(V2) \textbf{Deterministic Joint Density Sampling}: 
To maintain a fully deterministic algorithm, we pre-compute optimal deterministic samples from a higher-dimensional random vector that encompasses all \ac{CEM} iterations, \ie, we generate deterministic samples in dimension $\dimVarCem \cdot \numCEMIter$.
% \Ie, $[\xi^{(1)}_1, \ldots, \xi^{(1)}_{\numCEMIter}, \ldots, \xi^{(\numCEMSamples)}_1, \ldots, \xi^{(\numCEMSamples)}_{\numCEMIter}]$ are deterministically sampled from $\Gaussian(\vzero, \mI)$ in dimension $\dimVarCem \cdot \numCEMIter$.
% If the optimization variable has dimension $\dimVarCem$ and the \ac{CEM} runs for $\numCEMIter$ iterations, we generate samples in dimension $\dimVarCem \cdot \numCEMIter$. 
Each high-dimensional sample $[\tilde{\vVarCem}^{(i)\top}_1, \ldots, \tilde{\vVarCem}^{(i)\top}_{\numCEMIter}]\T$ is then partitioned into $\numCEMIter$ separate samples of dimension $\dimVarCem$, providing a unique, deterministic sample set for each iteration, which are then transformed using \cref{eq:lcd_sample_transform}.

(V3) \textbf{Combined Approach}: This hybrid scheme balances determinism within the optimization loop with stochastic exploration across time steps. A single set of high-dimensional samples is pre-computed as in (V2). At the beginning of each \ac{MPC} time step $k$, this set is rotated by a single random matrix $\mR_k \in SO(\dimVarCem)$. The resulting samples are then partitioned and used deterministically for all subsequent \ac{CEM} iterations within that time step.

\paragraph{Covariance Matrix Structures and Adaptation}
Given the sample set by one of the above schemes, the samples are transformed to match the current proposal distribution of the \ac{CEM} optimizer using \cref{eq:lcd_sample_transform}.
The use of deterministic samples is compatible with, \eg, maintaining a diagonal covariance matrix, or by adapting a full covariance matrix.
We focus on two approaches that incorporate temporal correlations, which are known to be beneficial for generating smooth control sequences~\cite{pinneriSampleefficientCrossentropyMethod2021}. 
The initial correlation structure is derived from the power spectral density of colored noise (\eg, pink noise, $\beta=1$) via the Wiener--Khinchin theorem, resulting in a Toeplitz structured correlation matrix $\mC_{\rho}$~\cite[pp. 576--578]{kay1993fundamentals}.

(M1) \textbf{Fixed Correlation with Adaptive Variance}: In this approach, the correlation structure of the proposal distribution is fixed throughout the optimization. 
The covariance matrix is constructed as $\mC_j = \diag(\vsigma_j) \mC_{\rho} \diag(\vsigma_j)$, where $\mC_{\rho}$ is the fixed time-correlation matrix and $\vsigma_j$ is the vector of marginal standard deviations. 
In each iteration, only the marginal variances are updated based on the elite set, which simplifies~\cref{eq:cem_cov_update} to 
% a per-dimension variance update
\begin{align}
    \vsigma^2_{j+1} & = \frac{1}{|\mathcal{E}_j|} \sum_{\vVarCem^{(i)} \in \mathcal{E}_j} (\vVarCem^{(i)} - \evVarCem_{j+1})^2 \enspace.
\end{align}
This reduces the number of parameters to be estimated and is less demanding on the number of elite samples; a minimum of two non-identical elite samples is sufficient.
The transformation in \cref{eq:lcd_sample_transform} then uses 
%\label{eq:lcd_fixed_corr}
$\mL_j = \diag(\vsigma_j) \mA_{\rho}$,
where $\mA_{\rho}$ is the matrix square root of $\mC_{\rho}$.

(M2) \textbf{Adaptive Full Covariance}: 
This strategy allows for greater flexibility by adapting the full covariance matrix $\mC_j$ in each iteration. 
The optimization is initialized with a structured covariance matrix based on colored noise as in the previous scheme. 
Subsequently, the entire matrix is updated using the maximum likelihood estimate from the elite set. 
This allows the optimizer to adapt the temporal correlations online. 
However, estimating a full $\dimVarCem \times \dimVarCem$ covariance matrix requires the elite set to contain at least $\dimVarCem + 1$ non-collinear samples~\cite{julierReducedSigmaPoint2002}.

The sample set variation schemes (V1--V3) and covariance adaptation methods (M1--M2) can be combined arbitrarily.
For instance, random rotations (V1) can be used with either a fixed correlation structure with adaptive variances (M1) or a full adaptive covariance matrix (M2).
This modularity allows the proposed deterministic sampling strategies to serve as drop-in replacements for the random sampling step in various \ac{CEM}--\ac{MPC} algorithms.
\section{Experiments}
\label{sec:evaluation}

To evaluate the proposed \ac{CEM}--\ac{MPC} with deterministic samples, we conduct experiments on two tasks: (i)~the mountain car~\cite{towers2024gymnasium} and (ii)~the cart-pole swing-up task~\cite{bartoNeuronlikeAdaptiveElements1983}.

In contrast to standard simulation environments, such as Gymnasium~\cite{towers2024gymnasium}, where the system dynamics are deterministic, we add additive Gaussian process noise when the optimized control inputs are applied to the system during simulation.
The simulated dynamics are then given by
\begin{align}
    \vx_{k+1} & = \va(\vx_k, u_k) + \vw_k \enspace,
\end{align}
where $\vw_k$ is sampled from a zero-mean Gaussian \ac{PDF} $\mathcal{N}(\vw; \vzero, \mC^\vw)$ with process noise covariance $\mC^\vw$.
However, the controller does not have access to the process noise or its \ac{PDF} and assumes deterministic dynamics $\va(\cdot, \cdot)$ for prediction, and therefore has to deal with random disturbances.
For both tasks, quadratic stage cost functions of the form $g_n(\vx_n, u_n) = (\vx_n - \vx_{\mathrm{g}})\T \mQ (\vx_n - \vx_{\mathrm{g}}) + r \cdot u_n^2$ are used, where $\vx_{\mathrm{g}}$ is the goal state, and $\mQ$ and $r$ are state and control weights, respectively. 
The terminal cost function is set to $g_{\horizon}(\vx_{k+\horizon}) = (\vx_{k+\horizon} - \vx_{\mathrm{g}})\T \mQ_\mat\horizon (\vx_{k+\horizon} - \vx_{\mathrm{g}})$.
The task-specific parameters are summarized in \cref{tab:task_parameters}.

\begin{table}[t]
    \centering
    \caption{Task-specific parameters for evaluation.}
    \label{tab:task_parameters}
    \setlength{\tabcolsep}{1pt}
    \begin{tabular}{lcc}
        \toprule
        Parameter & Mountain Car & Cart-Pole Swing-Up \\
        \midrule
        Control input limits & $u \in [-1, 1]$ & $u \in [-20, 20]\,\si{\newton}$ \\
        Goal state $\vx_{\mathrm{g}}$ ($\tilde{\vx}_{\mathrm{g}}$) & $[\nicefrac{\pi}{2}, 0]\T$ & $[0, 0, 1, 0, 0]\T$ \\
        State weights $\mQ$ & $\diag([1, 1])$ & $\diag([0.1, 0.1, 1, 0.1, 0.1])$ \\
        Control weight $r$ & $0.1$ & \num{1e-4} \\
        Terminal weights $\mQ_{\horizon}$ & $\diag([1, 1])$ & $\diag([10, 0.1, 10, 0.1, 0.1])$ \\
        Process noise $\mC^\vw$ & $\diag([0, \num{1e-7}])$ & $\diag([0, \num{1e-8}, 0, \num{1e-8}])$ \\
        Time discretization $\Delta t$ & \SI{3}{\second} & \SI{0.02}{\second} \\
        Total time steps $T$ & \num{150} & \num{300} \\
        Noise color for \ac{CEM} $\beta$ & $0.25$ & $1.0$ \\
        Initial $\evVarCem_{0}$ for \acs{CEM} & $\vzero$  & $\vzero$ \\
        Initial $\vsigma_0$ for \acs{CEM} & $\num{1.5}\cdot\vones$  & $\num{10}\cdot\vones$ \\
        \bottomrule
    \end{tabular}
\end{table}

For the evaluation, we denote the use of our deterministic sampling strategies as \emph{deterministic sampling CEM} (\acs{dsCEM}). 
Specifically, we refer to the adaptation methods from \cref{sec:cem_mpc_lcd_integration} as \emph{dsCEM-Var} (M1) and \emph{dsCEM-Cov} (M2), and for the variability schemes we add \emph{V1}, \emph{V2}, and \emph{V3}, when the specific scheme is used.
dsCEM-Var is evaluated with all three variability schemes, while dsCEM-Cov is only evaluated with scheme V3, to keep visualizations clear.

We compare our approach against the iCEM method~\cite{pinneriSampleefficientCrossentropyMethod2021}, where the only difference is that we swapped the random sampling step with our proposed deterministic sampling strategies, \ie, both methods have the same computational budget.
For a fair comparison, both methods share hyperparameters as used in~\cite{pinneriSampleefficientCrossentropyMethod2021}: a horizon $\horizon = 30$, $\numCEMIter = \num{3}$ iterations, momentum $\alpha = 0.1$, and shift-initialization warm-starting (see \cref{sec:cem_mpc:related_work}). 
The elite set size is $\numCEMElite = 10$ (or $\numCEMElite = 40$ for \ac{dsCEM}-Cov), and a fraction of the top \num{0.3} of the elite set is carried over to the next time step (rounded down).
The initial correlation for \ac{dsCEM} is derived from the noise color $\beta$ used in iCEM as described in~\cref{sec:cem_mpc_lcd_integration}.
To isolate the impact of the sampling strategy, iCEM's sample decay is disabled, ensuring an equal number of shooted trajectories for both methods.

We evaluate all methods over a sample size $\numCEMSamples$ ranging from \num{20} to \num{300}.
Each configuration is repeated for \num{100} runs with different random seeds.
The range for \ac{dsCEM}-Cov starts at \num{40} samples due to its higher elite sample requirement.
Additionally, we establish a performance baseline by running iCEM with an extensive sample size of \num{1e4}.

We evaluate controller performance using cumulative cost and control input smoothness.
The cumulative cost is the sum of stage costs $g_k(\vx_k, \vu_k)$ over the entire simulation.
The smoothness is measured using~\cite{powerLearningGeneralizableTrajectory2024}
\begin{align}
    S & = \sum_{k=1}^{T-1} \| \vu_{k} - \vu_{k-1} \|^2 \enspace,
\end{align}
where a lower value of $S$ indicates a smoother trajectory by penalizing large changes between consecutive control inputs.

\subsection{Mountain Car Task}
\label{sec:mountain_car}

The dynamics of the mountain car task~\cite{towers2024gymnasium} are given by
\begin{align}
    \begin{bmatrix}
        \dot{x}_1 \\
        \dot{x}_2
    \end{bmatrix}
    = \begin{bmatrix}
        x_2 \\
        - 0.0025 \cos(3 x_1)
\end{bmatrix} + \begin{bmatrix}
        0 \\
        0.0015
    \end{bmatrix} u \enspace,
\end{align}
where $x_1$ is the position and $x_2$ is the velocity. %, and $u$ is the control input.
The system is discretized using Runge--Kutta 4th order (RK4) integration. %with a time step of $\Delta t = \SI{3}{\second}$.
% We use a simulation time of $T = \num{150}$ and add process noise with covariance $\mC^\vw = \diag([0, \num{1e-7}])$ to the velocity.
The main challenge is that the underpowered car cannot drive directly up the hill and must perform a swing-up maneuver.
Unlike the standard task, our objective is to reach the goal state at the top of the hill \emph{and stop}, which is more difficult than reaching the top position with an arbitrary final velocity.
The state is initialized with position $x_1$ drawn uniformly from $[-0.7, -0.3]$, and with zero initial velocity.
% $\vx_{\mathrm{g}} = [0.6, 0]$ and stop, which is more difficult than only reaching the position with an arbitrary final velocity.
% The cost weights are $\mQ = \mQ_{\horizon} = \diag([1, 1])$ and $r = 0.1$.
% The iCEM noise color, used for our method's correlation initialization, is $\beta=0.25$.

\begin{figure}%
    \newcommand{\vspaceBelowSubfigure}{\vspace{4pt}}
    \centering%
    \begin{subfigure}{\columnwidth}
        \includegraphics[width=\linewidth]{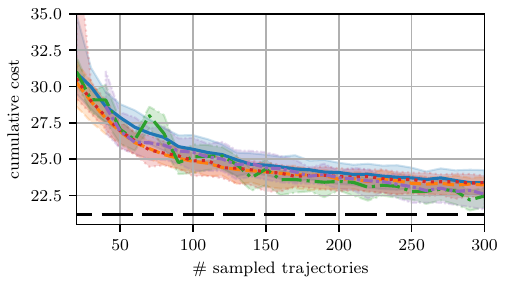}%
        \caption{Cumulative costs}%
        \label{fig:mountain_car_results:cum_cost}%
    \end{subfigure}
    \vspaceBelowSubfigure
    \begin{subfigure}{\columnwidth}
        \includegraphics[width=\linewidth]{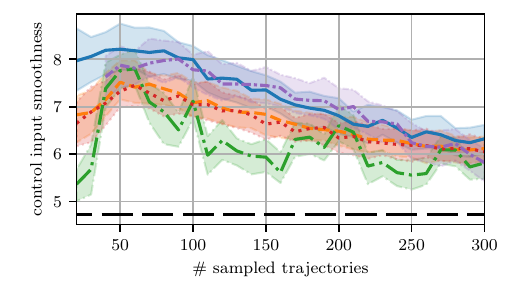}%
        \caption{Action smoothness}%
        \label{fig:mountain_car_results:action_smoothness}%
    \end{subfigure}
    \vspaceBelowSubfigure
    \begin{subfigure}{\columnwidth}
        \centering%
        \includegraphics[width=\linewidth]{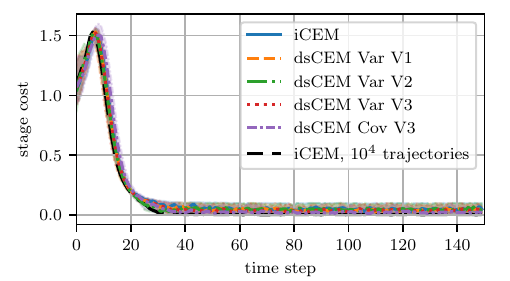}%
        \caption{Convergence (\num{50} control sequence samples)}%
        \label{fig:mountain_car_results:convergence}
    \end{subfigure}
    \caption{The results for the Mountain Car Task are given for \subref{fig:mountain_car_results:cum_cost} cumulative cost and \subref{fig:mountain_car_results:action_smoothness} control input smoothness over sample size, as well as for \subref{fig:mountain_car_results:convergence} convergence behavior for a fixed sample size of \num{50}. All plots show the median (line) and the interquartile range (shaded area) across \num{100} runs.
    The different methods' colors are consistent across all plots.
    }%
    \label{fig:mountain_car_results}%
\end{figure}%

The results for the mountain car task are presented in~\cref{fig:mountain_car_results}.
In terms of cumulative cost (\cref{fig:mountain_car_results:cum_cost}), all proposed \ac{dsCEM} methods outperform iCEM, particularly at small sample sizes.
While the performance of all methods converges for larger sample sizes, they do not reach the baseline established by iCEM with \num{1e4} samples.
Regarding smoothness (\cref{fig:mountain_car_results:action_smoothness}), all \ac{dsCEM} variants, except for \ac{dsCEM}-Cov V3, produce smoother control trajectories than iCEM.
Notably, \ac{dsCEM}-Var V2 consistently achieves the highest degree of smoothness (lowest values) across most sample sizes.
An interesting trend is visible in the smoothness measure, where the score for all methods initially increases, peaking around a sample size of \num{50}, before decreasing again.
This peak likely indicates the sample budget required to discover aggressive swing-up maneuvers that are effective for reducing costs.
With fewer samples, the optimizer settles for smoother but less optimal trajectories, while larger budgets allow it to find and subsequently refine these more dynamic strategies.
The convergence behavior for a sample size of \num{50} is depicted in~\cref{fig:mountain_car_results:convergence}.
While the median stage cost evolves similarly for all methods, our proposed \ac{dsCEM} variants demonstrate slightly faster convergence, particularly around time step \num{30}, and their performance closely approaches the extensive iCEM baseline.
As the costs approach zero, the differences between the methods become less significant because all controllers primarily focus on counteracting disturbances.

\subsection{Cart-Pole Swing-Up Task}
\label{sec:cart_pole}

\newcommand{\stateX}[1]{\ensuremath{\mathsf{x}_{#1}}}

The dynamics of the frictionless cart-pole swing-up task are given by~\cite{bartoNeuronlikeAdaptiveElements1983}
\begin{align}%
    \ddot{\phi} & = 
        \frac{g \sin(\phi) - \cos(\phi) \cdot 
            \frac{
                u + m_{\mathrm{p}} l \dot{\phi}^2 \sin(\phi) 
            }        
            {  
                m_{\mathrm{p}} + m_{\mathrm{c}}
            }
        }
        {
            l \left(\frac{4}{3} - \frac{m_{\mathrm{p}} \cos^2(\phi)}{ m_{\mathrm{p}} + m_{\mathrm{c}}}\right) 
        }
    \enspace,
    \\
    \ddot{\stateX{}} & = 
        \frac{
            u + m_{\mathrm{p}} l \left(\dot{\phi}^2 \sin(\phi) - \ddot{\phi} \cos(\phi) \right)}
        {
            m_{\mathrm{p}} + m_{\mathrm{c}}
        }
    \enspace,
\end{align}
where the state is $\vx = [\stateX{}, \dot{\stateX{}}, \phi, \dot{\phi}]\T$, with cart position $\stateX{}$, pole angle $\phi$ (where $\phi=0$ is the upright position), and their respective (angular) velocities. 
% The control input is the force $u \in [-10, 10]\,\si{\newton}$ applied to the cart, and 
The physical parameters are cart mass $m_{\mathrm{c}} = \SI{1}{\kilo\gram}$, pole mass $m_{\mathrm{p}} = \SI{0.1}{\kilo\gram}$, pole length $l = \SI{0.5}{\meter}$, and gravity $g = \SI{9.81}{\meter\per\second\squared}$.
The system is discretized using RK4. 
% with a time step of $\Delta t = \SI{0.02}{\second}$ and simulated for $T = \num{300}$ steps, with process noise of covariance $\mC^\vw = \diag([0, \num{1e-8}, 0, \num{1e-8}])$ added to the velocities.
% where $x_1$ is the cart position (in \si{\meter}), $x_2$ the cart velocity (in \si{\meter\per\second}), $x_3$ the pole angle (in \si{\radian}), $x_4$ the pole angular velocity (in \si{\radian\per\second}), and $u \in [-10, 10] \si{\newton}$ the control input (force applied to the cart). 
%
The initial state is set to zero for the cart's position and both velocities, while the pole's angle $\phi$ is uniformly sampled from $[\ang{145}, \ang{215}]$.
The costs are evaluated using an augmented state vector $\tilde{\vx} = [x_1, x_2, \cos(x_3), \sin(x_3), x_4]$ to account for the periodicity of the angle $x_3$.
% As in the mountain car task, stage cost is quadratic and penalizes deviations from the goal state $\tilde{\vx}_{\mathrm{g}} = [0, 0, 1, 0, 0]\T$ and control effort, with weights $\mQ = \diag([0.1, 0.1, 1, 0.1, 0.1])$ and $R = \num{1e-4}$.
% The quadratic terminal cost is weighted by $\mQ_{\horizon} = \diag([10, 0.1, 10, 0.1, 0.1])$.
% The used noise color is $\beta = 1.0$.

\begin{figure}%
    \centering%
    \newcommand{\vspaceBelowSubfigure}{\vspace{1.5pt}}
    \begin{subfigure}{\columnwidth}
        \includegraphics[width=\linewidth]{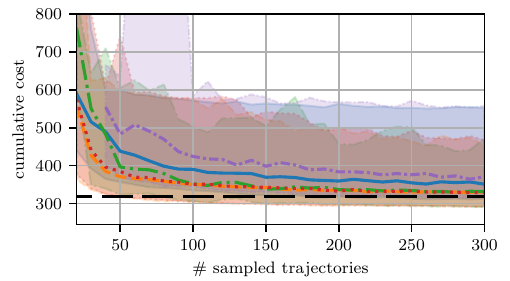}%
        \caption{Cumulative costs}%
        \label{fig:cart_pole:cum_cost}%
    \end{subfigure}
    \begin{subfigure}{\columnwidth}
        \includegraphics[width=\linewidth]{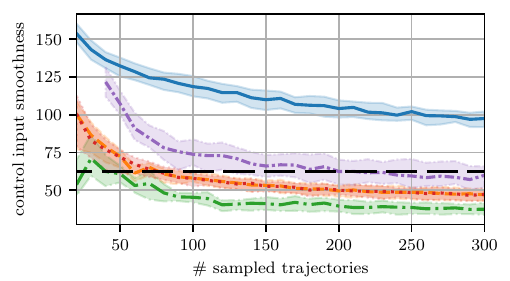}%
        \caption{Action smoothness}%
        \label{fig:cart_pole:action_smoothness}%
    \end{subfigure}
    \vspaceBelowSubfigure
    \begin{subfigure}{\columnwidth}
        \centering%
        \includegraphics[width=\linewidth]{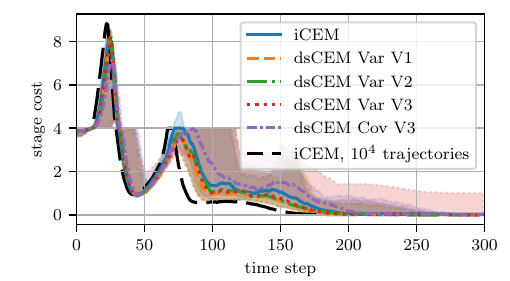}%
        \caption{Convergence (\num{50} control input samples)}%
        \label{fig:cart_pole:convergence}
    \end{subfigure}
    \vspaceBelowSubfigure
    \begin{subfigure}{\columnwidth}
        \centering%
        \includegraphics[width=\linewidth]{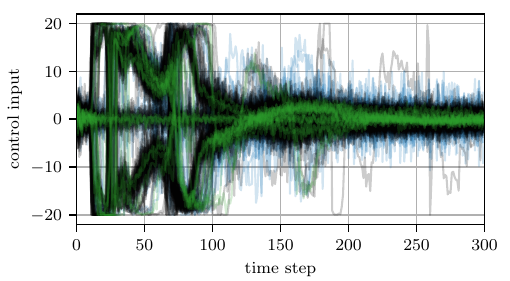}%
        \caption{Applied control inputs for \num{20} randomly selected runs, with each method distinguished by color. The optimization for each run was performed using \num{300} samples.}%
        \label{fig:cart_pole:action_time_series}
    \end{subfigure}
    \caption{Results for the cart-pole task. The different methods' colors are consistent across all plots.
        To give intuition about runtime, for \num{100} samples iCEM and \ac{dsCEM} both take \SI{166}{\milli\second} per \ac{MPC} step on one Intel Xeon Platinum 8358 core, indicating no measurable overhead from deterministic sampling and variation schemes; parallelization can further reduce runtime.
    }
    \label{fig:cart_pole_results}%
\end{figure}%

% The results for the cart-pole swing-up task are presented in \cref{fig:cart_pole_results}.
In terms of cumulative cost (\cref{fig:cart_pole:cum_cost}), most proposed \ac{dsCEM} methods outperform iCEM and closely approach the performance of the extensive iCEM baseline.
However, the submethod based on the full covariance (\ac{dsCEM} Cov V3) struggled to match the performance of our variance-based methods (\ac{dsCEM} Var V1--V3).
This suggests that using a high-dimensional covariance matrix may not be suitable for this task with small sample sizes.
Regarding control input smoothness (\cref{fig:cart_pole:action_smoothness}), all \ac{dsCEM} methods demonstrate superior performance over iCEM, with \ac{dsCEM}-Var V2 being the smoothest across all sample sizes.
Notably, with more than \num{50} samples, \ac{dsCEM}-Var V2 even surpasses the smoothness of the extensive iCEM baseline.
This is further illustrated in \cref{fig:cart_pole:action_time_series}, which shows that the control input trajectories of \ac{dsCEM}-Var V2 are significantly less noisy than those of iCEM with the same sample budget, and even smoother than the extensive baseline.
The convergence plot for a sample size of \num{50} (\cref{fig:cart_pole:convergence}) confirms that most \ac{dsCEM} variants converge faster than iCEM.
However, the wide interquartile range for \ac{dsCEM}-Var V3 suggests that this specific variant is less robust.
% In terms of computational cost (\num{100} samples), iCEM and \ac{dsCEM} both take \SI{166}{\milli\second} per MPC step on one Xeon 8358 core, indicating no measurable overhead from deterministic sampling and variation schemes. %; parallelization can further reduce runtime.

\subsection{Discussion}
\label{sec:experiments:discussion}

The experimental results demonstrate that replacing standard random sampling in \ac{CEM} with our proposed deterministic sampling strategies, termed \ac{dsCEM}, leads to significant improvements in performance and sample efficiency. 
Across both the mountain car and cart-pole swing-up tasks, the \ac{dsCEM} variants consistently outperform the baseline iCEM method, particularly when the number of samples is limited.

A key finding is the superior control input smoothness achieved by the \ac{dsCEM} methods. 
While smooth controls can also be encouraged by adding explicit smoothness penalties, this introduces additional hyperparameters to trade off control performance and smoothness; in our view, the optimizer should produce smooth controls without such extra tuning.
This is especially pronounced for the fully deterministic variant \ac{dsCEM}-Var~V2, which consistently yields the smoothest control trajectories. 
This outcome is expected, as deterministic sample sets provide low-discrepancy coverage of the sampling space, mitigating the clustering and gaps inherent to random sampling. 
Notably, in the cart-pole swing-up task, \ac{dsCEM} produces smoother control inputs than even the extensive iCEM baseline, despite using a fraction of the samples. 
This highlights the ability of deterministic sampling to find high-quality, smooth solutions with remarkable efficiency.

In terms of cumulative cost, \ac{dsCEM} shows a clear advantage at lower sample sizes, indicating faster convergence to effective control strategies.
Although all methods tend to converge as sample sizes increase, the gains from \ac{dsCEM} in the low-sample regime are critical for practical applications.
% However, the \ac{dsCEM}-Cov method struggled to match the performance of the other methods on the cart-pole task.
% This suggests that using a high-dimensional covariance matrix may not be suitable for all tasks with small sample sizes.

The enhanced sample efficiency of \ac{dsCEM} has profound implications for the deployment of \ac{MPC}.
Reducing the required number of samples lowers computational demand, a critical factor for real-time performance.
This reduction impacts four key stages of the \ac{CEM} algorithm:
(i)~the creation of random or deterministic samples;
(ii)~the determination of the elite set;
(iii)~the computationally expensive trajectory shooting, especially for complex system dynamics; and
(iv)~the evaluation of cost functions for each trajectory and all its stages.
Furthermore, as hardware generally has limited parallelization capabilities, even a slight reduction in sample size can prevent substantial computational overhead.
For example, on such hardware, an algorithm requiring one more sample than the hardware's capacity can \emph{double the execution time}.
By achieving superior performance with fewer samples, \ac{dsCEM} promises to apply \ac{MPC} to systems with more complex models and longer prediction horizons without sacrificing real-time capability.
\section{CONCLUSION}
\label{sec:conclusion}

In this paper, we addressed the limitations of random sampling in \ac{CEM}--\ac{MPC} by introducing \ac{dsCEM}, a novel framework that leverages deterministic samples based on \acp{LCD}. 
We proposed several schemes for integrating these samples, demonstrating that this approach is a modular, drop-in replacement for the conventional sampling step.
Our experimental evaluation on two nonlinear control benchmarks showed that \ac{dsCEM} consistently outperforms the state-of-the-art iCEM method in terms of both cumulative cost and control input smoothness, especially in the critical low-sample regime.
% Notably, our method achieved superior action smoothness with lower computational budget, even surpassing an extensive iCEM baseline using \num{1e4} samples.
%
% These findings highlight the potential of deterministic sampling to significantly enhance the sample efficiency of \ac{CEM}--\ac{MPC}, making it a more viable option for real-time control on computationally constrained hardware.
These findings highlight the potential of deterministic sampling to significantly improve the efficiency of \ac{CEM}--\ac{MPC}, making it a promising option for real-time control on computationally constrained hardware.

As our approach is orthogonal to learning-based improvements, a viable option is to combine \ac{dsCEM} with these techniques, \eg, using a learned policy to warm-start the optimization process and apply it to robotic systems.
% As our approach is orthogonal to learning-based improvements, future work will focus on combining \ac{dsCEM} with such techniques, for instance by using a learned policy to warm-start the optimization.
% Investigating the scalability of this combined approach to higher-dimensional robotic systems also remains a promising direction.

% \input{sections/04_1_uce.tex}

\balance
% \IEEEtriggeratref{4} % This is better than balance when IEEEtran is used. Set the number manually!

%%%%%%%%%%%%%%%%%%%%%%%%%%%%%%%%%%%%%%%%%%%%%%%%%%%%%%%%%%%%%%%%%%%%%%%%%%%%%%%%
% appendix
%%%%%%%%%%%%%%%%%%%%%%%%%%%%%%%%%%%%%%%%%%%%%%%%%%%%%%%%%%%%%%%%%%%%%%%%%%%%%%%%
% \input{sections/99_appendix}

%%%%%%%%%%%%%%%%%%%%%%%%%%%%%%%%%%%%%%%%%%%%%%%%%%%%%%%%%%%%%%%%%%%%%%%%%%%%%%%%

%%%%%%%%%%%%%%%%%%%%%%%%%%%%%%%%%%%%%%%%%%%%%%%%%%%%%%%%%%%%%%%%%%%%%%%%%%%%%%%%
% bibliography

% \bibliographystyle{bib/IEEEtran_Capitalize}
%\bibliographystyle{IEEEtran_Capitalize_less_authors} % use et al.
% \bibliography{bib/literature}
%\bibliography{literature.bib}

\bibliographystyle{IEEEtran}
\bibliography{IEEEabrv,bib/literature}

%%%%%%%%%%%%%%%%%%%%%%%%%%%%%%%%%%%%%%%%%%%%%%%%%%%%%%%%%%%%%%%%%%%%%%%%%%%%%%%%

\end{document}